\documentclass[epjCONF,columns]{svjour} 
\usepackage[square,numbers]{natbib}
\usepackage{graphics}
\usepackage[varg]{txfonts} 
\usepackage[latin1]{inputenc}

\newcommand{\BV}{Brunt-V\"ais\"al\"a\ }

\session-title{Conference Title, to be filled}
\begin{document}
\title{Gravito-inertial modes in a differentially rotating spherical shell}
\author{Giovanni M. Mirouh\inst{1,2}\fnmsep\thanks{\email{giovanni.mirouh@irap.omp.eu}} 
\and Cl\'ement Baruteau\inst{1,2} \and Michel Rieutord\inst{1,2} \and J\'er\^ome Ballot\inst{1,2}} 
\institute{Universit\'e de Toulouse, UPS-OMP, IRAP, Toulouse, France
 \and CNRS, IRAP, 14 avenue Edouard Belin, 31400 Toulouse, France}
\abstract{
While many intermediate- and high-mass main sequence stars are rapidly and differentially rotating, the effects of rotation on oscillation modes are poorly known. 
In this communication we present a first study of axisymmetric gravito-inertial modes in the radiative zone of a differentially rotating star. 
We consider a simplified model where the radiative zone of the star is a linearly stratified rotating fluid within a spherical shell, 
with differential rotation due to baroclinic effects. 
We solve the eigenvalue problem with high-resolution spectral computations and determine the propagation domain of the waves through 
the theory of characteristics. We explore the propagation properties of two kinds of modes: 
those that can propagate in the entire shell and those that are restricted to a subdomain. 
Some of the modes that we find concentrate kinetic energy around short-period shear layers known as attractors. 
We describe various geometries for the propagation domains, conditioning the surface visibility of the corresponding modes.}

\maketitle
\section{Introduction}
\label{intro}
Gravito-inertial modes are low-frequency modes restored by buoyancy and Coriolis forces. In rotating stars they may be excited by internal mechanisms, 
such as the $\kappa$-mechanism, or for planet-harbouring stars by tidal effects. 
Determining how tidally-excited waves deposit their energy and angular momentum helps predicting the orbital evolution of close-in planets.
Inertial modes in a differentially rotating convective layer have been studied \cite{BR13} as well as gravito-inertial modes in a 
radiative shell with solid-body rotation \cite{DRV99}, but the influence of differential rotation on gravito-inertial modes in a radiative zone is still unknown.

We wish to have more insights into the mode properties of high- and intermediate-mass main sequence stars, such as $\delta$ Scuti, Slowly Pulsating B, 
and $\beta$ Cephei stars. 
High rotation rates have been detected in most of these stars \cite{royer09}, while the radial differential rotation increases throughout their evolution \cite{ELR13}.
The oscillation properties of rapidly and differentially rotating stars are less constrained than those of slow rotators. 
For instance, regular patterns are hardly found in $\delta$ Scuti spectra \cite{mirouh_etal14a}.
The fundamental parameters are also modified by rotation, and may explain stars detected outside of their expected instability domain \cite{salmon14}.

Gravito-inertial modes probe the internal layers of the stellar radiative zone, around the convective core. 
A better characterisation of the oscillations will help constrain the physical model, in particular the core size.
\section{Model}
\label{sec:model}
We consider a viscous fluid enclosed in a spherical shell.
We impose a linear background temperature gradient through the radiative shell, resulting in a linear stable stratification.
The \BV frequency then reads $n(r) = N \times r$ with $N$ a constant \cite[see][for more details]{mirouh_etal14b}. 
According to \cite{HR14}, with no-slip boundary conditions on both sides, this leads to the following shellular differential rotation profile $\Omega(r)$:
\begin{equation}
  {{\Omega(r)}\over{\Omega(R)}} = 1 + \int\limits^R_r{\frac{n^2(r')}{r'}} dr' = 1 + \frac{N^2}{2} \left(1-\frac{r^2}{R^2}\right).
\end{equation}
We also make use of the Boussinesq approximation. We consider a radiative zone going from $r= \eta R = 0.35 R$ to $r =R$, $R$ being the stellar radius.

We determine the properties of the linear oscillations in this model using two methods \cite[][and references therein]{BR13, DRV99}. 
\begin{enumerate}
  \item 
We solve the eigenvalue problem of the stellar oscillations, namely the linearised equations of motion, energy, and mass conservation, 
including all dissipative terms.
                  
  \item
We compute the paths of characteristics of the associated adiabatic case \cite[e.g.][]{FS82a}. 
\end{enumerate}

The dissipative properties of the fluid are characterised by the Prandtl and the Ekman numbers, which are respectively defined as:
\begin{equation}
  {\rm Pr} = \frac{\nu}{\kappa},\qquad {\rm E} = \frac{\nu}{\Omega(R) R^2},
\end{equation}
where $\nu$ is the kinematic viscosity and $\kappa$ the thermal diffusivity of the fluid.

In stars, $\rm{Pr} \sim 10^{-5}$ and $\rm{E} \sim 10^{-10} - 10^{-16}$. However, given practical resolution limitations \cite{VRBF07},
we set $\rm{Pr} = 10^{-2}$ and $\rm{E} = 10^{-8}$ for a first numerical exploration.
Varying E and Pr parameters changes the minimum length-scale at which the energy of a mode may be focused along the attractor of characteristics.

\section{Mode classification}
Depending on the parameters and the position in the star, the solutions of the adiabatic problem may be evanescent or oscillatory. 
Therefore, for a given set of \BV frequency $N^2$ and wave frequency $\omega$, eigenmodes may occupy only a fraction of the spherical shell. 
This permits a classification of the modes, depending on whether the mode can propagate in the whole shell or in only part of it.
\begin{figure}
  \resizebox{\hsize}{!}{\includegraphics*{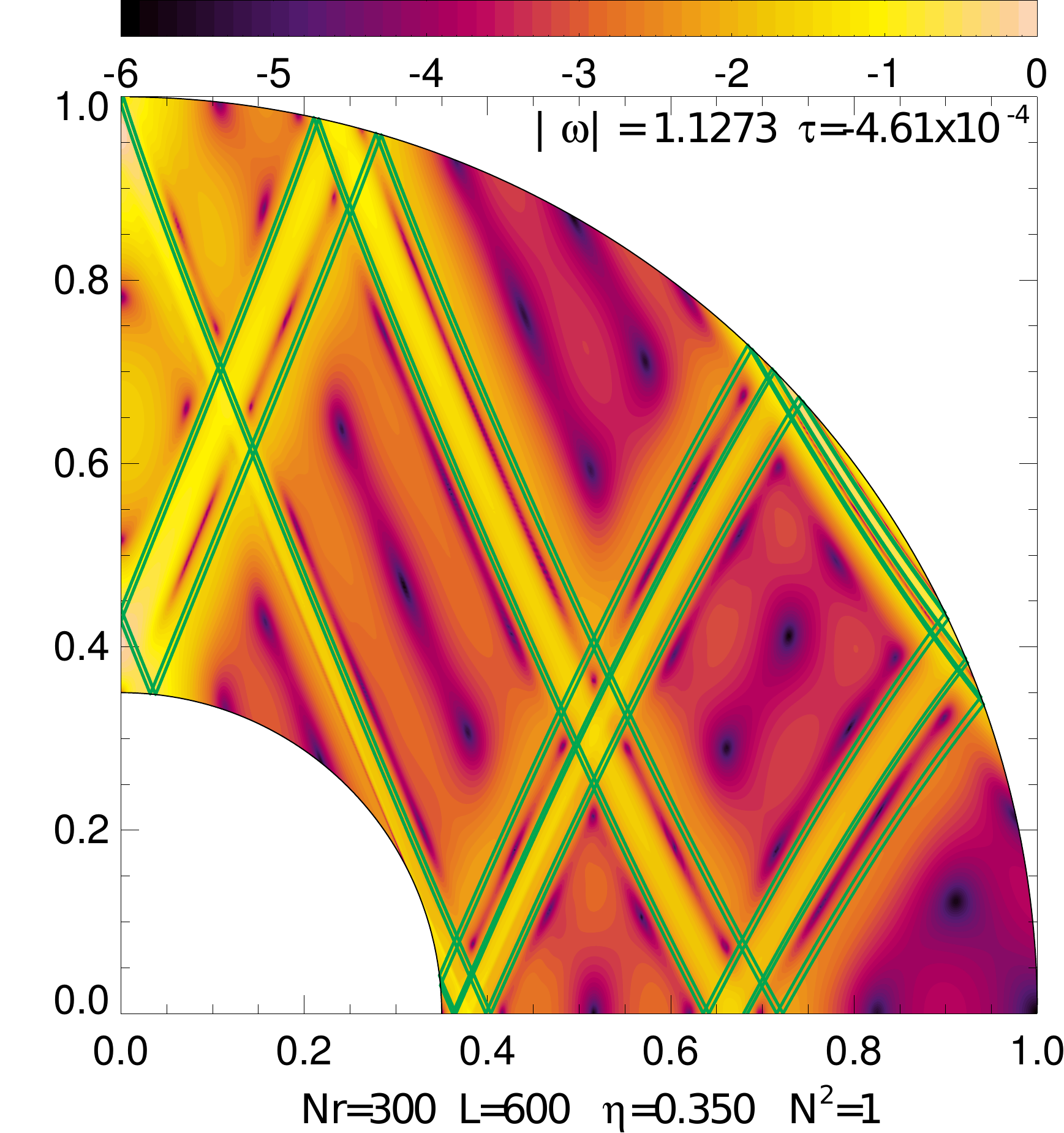}  \includegraphics*{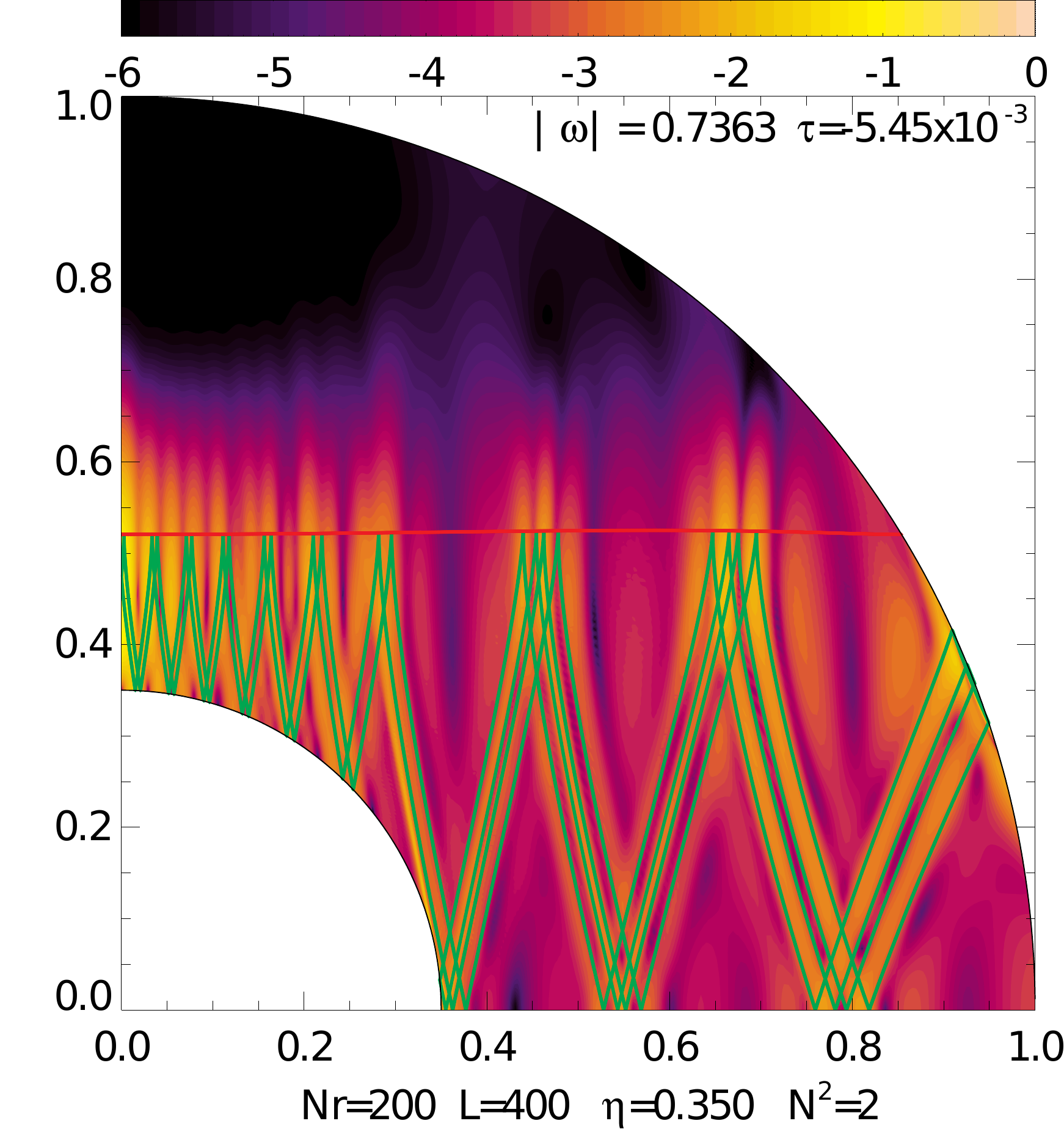}} \\
  \resizebox{\hsize}{!}{\includegraphics*{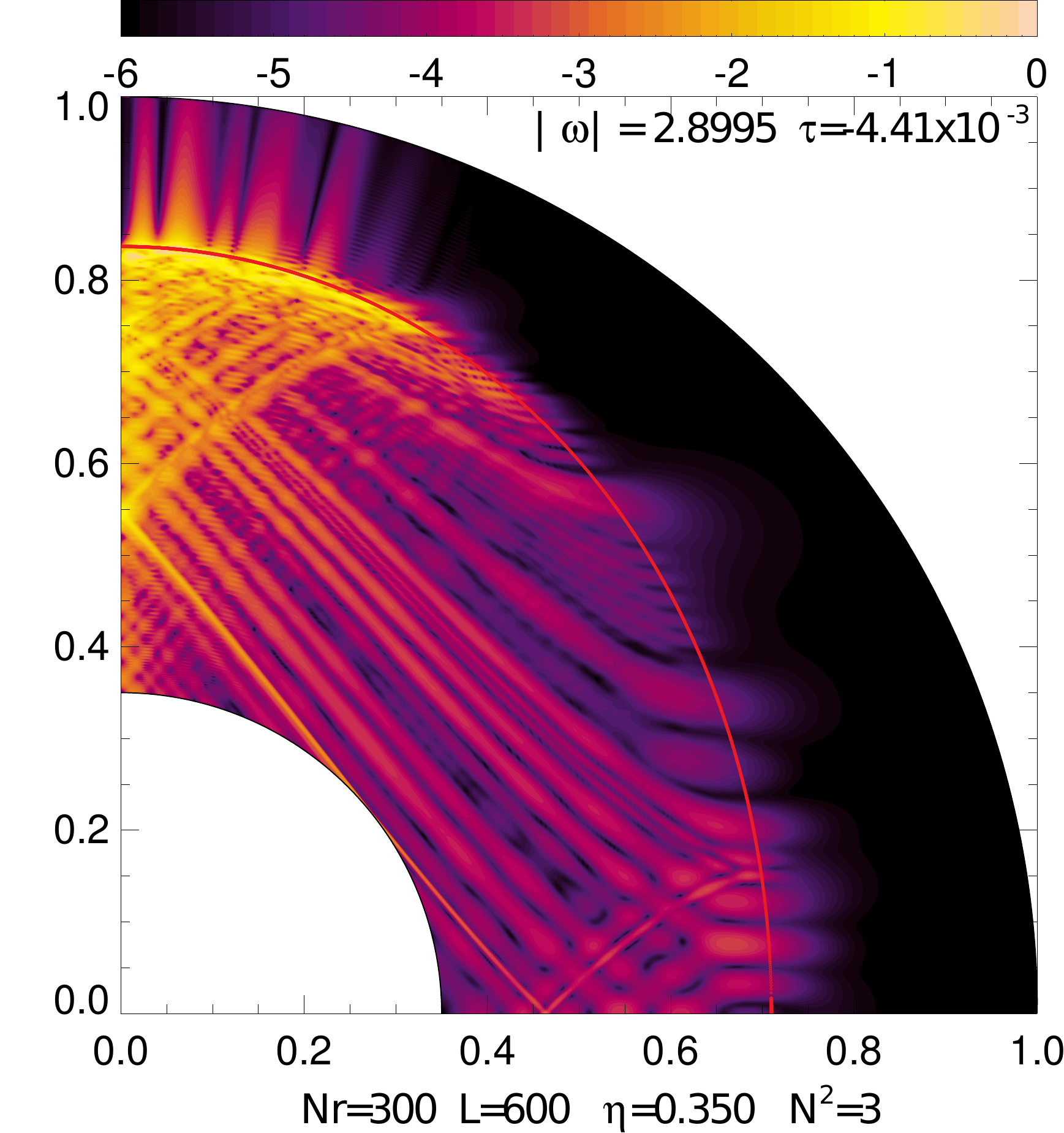}  \includegraphics*{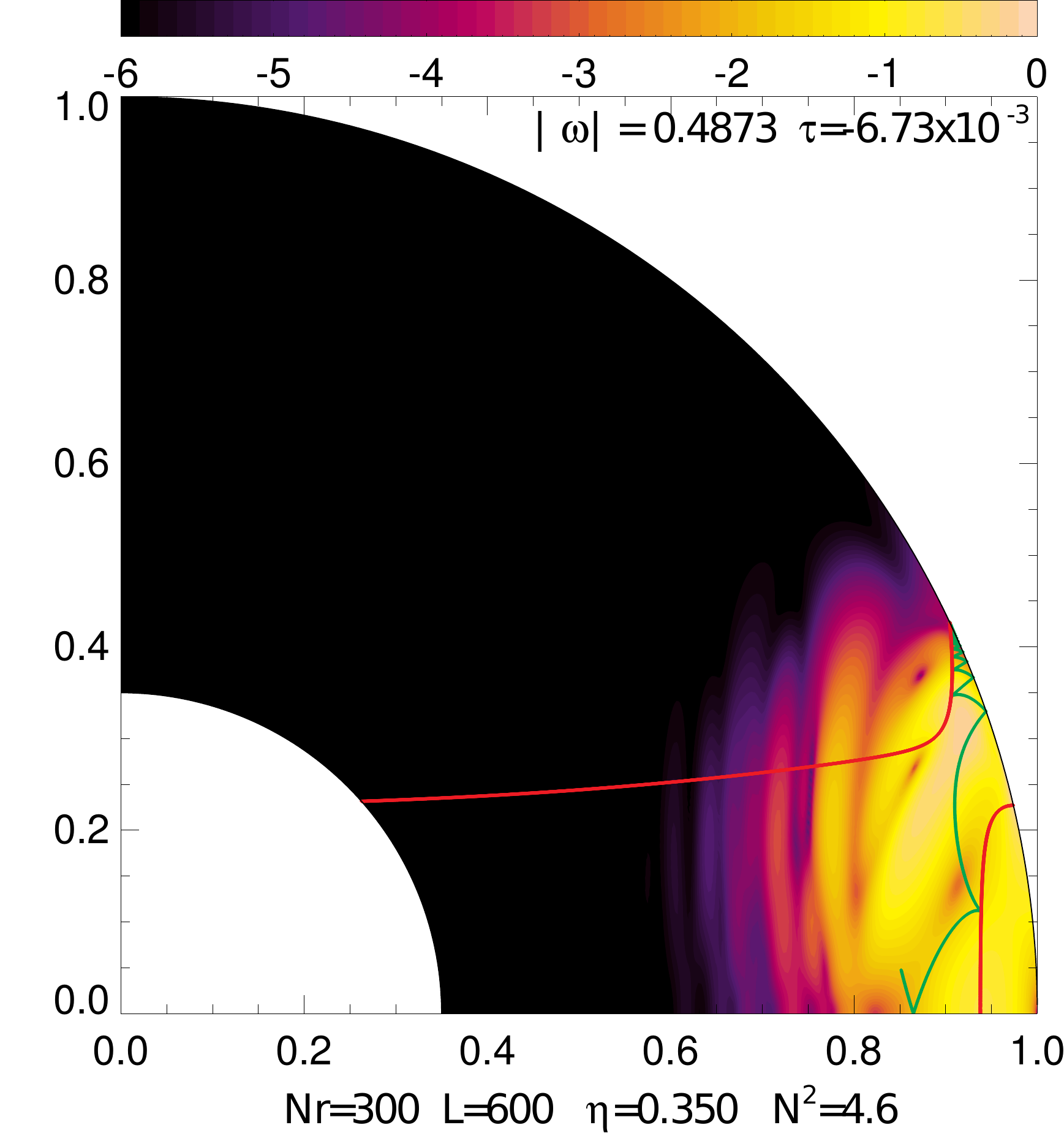}}
  \caption{
  Meridional slices of kinetic energy obtained by solving the dissipative linearised hydrodynamics equations,
  with attractors (green) and turning surfaces (red) overplotted. The energy is plotted on a logarithmic scale and normalised to its maximum value.
  The various geometries are discussed in the main text.
  }
  \label{fig:modes}
\end{figure}

Figure \ref{fig:modes} shows the kinetic energy distribution of some axisymmetric ($m=0$) modes for various values of $N^2$ and $\omega$, in a meridional plane.
The energy is focused along shear layers, which correspond to the trajectory of characteristics computed in the non-diffusive case.
The top left mode spans the whole shell, and the induced surface temperature perturbation might be detected by observations. 
The top right mode shows a turning surface: the propagation domain is limited to a subdomain of the shell. 
For this mode, the temperature perturbation only reaches the surface around the equator, making the oscillation visible only for large inclinations of the rotation
axis on the line of sight.
The bottom left mode also has a turning surface, but the oscillations are confined into an inner subdomain that does not reach the surface. This mode should
not be visible in lightcurves. Notice how the kinetic energy is spread through the subdomain with no convergence on a short-period attractor.
Finally, the bottom right panel shows an oscillation trapped in a wedge. The wedge is an acute angle created by turning surfaces and shell boundaries. 
It results in a focusing of the kinetic energy in the wedge, and impacts the visibility and dissipation properties of the mode. The occurence and properties of 
wedge trapped modes must be studied more thoroughly.

We have classified oscillation modes into different categories in a $(N^2, \omega)$ plane according to their propagation properties. 
Further details will be presented in a forthcoming paper.\\ 
We also compute the Lyapunov exponents, quantifying the convergence of the characteristics towards a short-period attractor. The faster the convergence, the more 
damped the mode and the less likely it is expected to be observed \cite{mirouh_etal14b}.

\section{Conclusions and future prospects}
For the first time, we compute the oscillations of a differentially rotating radiative region of star, where differential rotation is part of the 
baroclinic flow triggered by the combined effects of rotation and stable stratification. 

We have given a first view of axisymmetric eigenmodes which may propagate in such a background flow. The next steps include investigating non-axisymmetric modes
and the use of more realistic \BV frequency profiles, 
before tackling more realistic configurations with two-dim-ensional compressible stellar models, as the ESTER models \cite{ELR13}.

\bibliography{bibnew}{}

\begin{thebibliography}{10}

\bibitem{BR13}
C.~{Baruteau}, M.~{Rieutord}, J. Fluid Mech. \textbf{719}, 47 (2013)

\bibitem{DRV99}
B.~Dintrans, M.~Rieutord, L.~Valdettaro, J. Fluid Mech. \textbf{398}, 271
  (1999)

\bibitem{royer09}
F.~{Royer}, \emph{{On the Rotation of A-Type Stars}}, in \emph{The Rotation of
  Sun and Stars}, edited by J.P. {Rozelot}, C.~{Neiner} (2009), Vol. 765 of
  \emph{Lecture Notes in Physics, Berlin Springer Verlag}, pp. 207--230

\bibitem{ELR13}
F.~{Espinosa Lara}, M.~{Rieutord}, A\&A \textbf{552}, A35 (2013)

\bibitem{mirouh_etal14a}
G.M. {Mirouh}, D.R. {Reese}, F.~{Espinosa Lara}, J.~{Ballot}, M.~{Rieutord},
  \emph{{Asteroseismology of fast-rotating stars: the example of {$\alpha$}
  Ophiuchi}}, in \emph{IAU Symposium}, edited by J.A. {Guzik}, W.J. {Chaplin},
  G.~{Handler}, A.~{Pigulski} (2014), Vol. 301 of \emph{IAU Symposium}, pp.
  455--456

\bibitem{salmon14}
S.J.A.J. {Salmon}, J.~{Montalb{\'a}n}, D.R. {Reese}, M.A. {Dupret},
  P.~{Eggenberger}, A\&A \textbf{569}, A18 (2014)

\bibitem{mirouh_etal14b}
G.M. {Mirouh}, C.~{Baruteau}, M.~{Rieutord}, J.~{Ballot},
  \emph{{Gravito-inertial modes in a differentially rotating spherical shell}},
  in \emph{SF2A-2014: Proceedings of the Annual meeting of the French Society
  of Astronomy and Astrophysics}, edited by J.~{Ballet}, F.~{Bournaud},
  F.~{Martins}, R.~{Monier}, C.~{Reylé} (2014)

\bibitem{HR14}
D.~Hypolite, M.~Rieutord, to appear in A\&A \textbf{1}, 1 (2014)

\bibitem{FS82a}
S.~{Friedlander}, W.~{Siegmann}, J. Fluid Mech. \textbf{114}, 123 (1982a)

\bibitem{VRBF07}
L.~Valdettaro, M.~Rieutord, T.~Braconnier, V.~Fraysse, J. Comput. and Applied
  Math. \textbf{205}, 382 (2007), \texttt{physics/0604219}

\end{thebibliography}
\bibliographystyle{epj}

\end{document}